\UseRawInputEncoding 
\documentclass[aps,prb,amsmath,amssymb,preprint,longbibliography, superscriptaddress,showkeys]{revtex4-1}
\usepackage{graphicx}
\usepackage{dcolumn}
\usepackage{bm}
\usepackage{mathrsfs}
\usepackage{subfigure}
\usepackage{natbib}
\usepackage{amssymb}
\usepackage{multirow}
\usepackage{float}
\usepackage{color}
\usepackage{bbm}

\begin{document}

\title{Multiband Superconductors: Two Characteristic Lengths for Each Contributing Condensate}

\author{Yajiang Chen}
\affiliation{Key Laboratory of Optical Field Manipulation of Zhejiang Province, Department of
Physics, Zhejiang Sci-Tech University, 310018 Hangzhou, China}

\author{A. A. Shanenko}
\affiliation{HSE University, 101000 Moscow, Russia}

\begin{abstract}
Traditionally, the characteristic length of a superconducting condensate is associated with the spatial distribution of the corresponding gap function. However, the superconducting condensate is the quantum condensate of Cooper pairs and thus, the broader readership is more familiar with the concept of the Cooper-pair wave function. For conventional single-band superconductors, the gap function coincides with the center-of-mass wave function of a Cooper pair up to the coupling constant, and the corresponding gap- and wave-function characteristic lengths are the same. Surprisingly, we find that in two-band superconductors, these lengths are the same only near the critical temperature. At lower temperatures they can significantly deviate from each other, and the question arises as to which of these lengths should be given the preference when specifying the spatial scale of the band-dependent condensate in multiband superconducting materials.  
\end{abstract}

\maketitle

\section{Introduction}

When multiple condensates coexist in one system, they can interfere constructively or destructively. This can result in unconventional coherent behavior not present in single-condensate systems~\cite{MP2015, Lin2014, Tanaka2015}. We can mention the multiband crossover from the Bardeen-Cooper-Schrieffer (BCS) regime of loosely bound pairs to the bosonic condensate of molecule-like pairs~\cite{Lubashevsky2012,Okazaki2014, Chen2012}, possible fractional vortices~\cite{Babaev2002, Bluhm2006,Lin2013}, chiral solitons~\cite{Tanaka2001,Tanaka2015}, an enhancement of the intertype superconductivity~\cite{Vagov2016}, the multiband screening of the pair fluctuations~\cite{Salasnich2019,  Saraiva2020,  Saraiva2021, Shanenko2022} etc. Such phenomena are pronounced when the spatial profiles (spatial scales) of contributing condensates are different. The perception that the physics of systems with multiple condensates must be more complex than the physics of a single condensate is so appealing that there is even a long-discussed idea of a special type of superconductivity for multiband superconductors~\cite{Moshchalkov2009,Kogan2011, Babaev2012, Kogan2012}.

When speaking about the spatial scale of a band-dependent superconducting condensate, one usually means the characteristic length $\xi_{\Delta j}$~(the coherence length) of the gap function $\Delta_j({\bf r})$, where $j$ enumerates overlapping bands. Experimentally, $\xi_{\Delta j}$ can be extracted from the muon spin rotation measurements~\cite{Callaghan2005}, from the scanning tunnelling microscopy (STM) results~\cite{Fente2016, Fente2018}, from the thermal conductivity of the vortex state~\cite{Boaknin2003}, or from the upper critical field~\cite{Gennes1966}. However, one should keep in mind that relations between the gap-function distribution and the experimentally measured characteristics are generally model dependent, which complicates the interpretation of the experimental results. For example, the STM probes the local density of states (DOS) connected with the quasiparticle spatial distribution. The local relation between the zero-energy local DOS and the gap function is well-known only for the diffusive regime near the upper critical magnetic field~\cite{Gennes1964}. At lower fields the situation is significantly more complicated, especially in multiband superconducting materials~\cite{Nakai2002,Koshelev2003,Vargunin2019}. 

Different low/high field dependence of the vortex-core size in the two-band material NbSe$_2$ demonstrates~\cite{Callaghan2005} that there are two kinds of localized (in-gap) quasiparticles, which confirms the existence of two condensate lengths $\xi_{\Delta 1}$ and $\xi_{\Delta 2}$, since the vortex core is mainly determined by the in-gap quasiparticles~\cite{Caroli1964, Gygi1991}. However, recent STM results demonstrate that the vortex core of the two-band superconductor CaKFe$_4$As$_4$~\cite{Fente2018} exhibits only one characteristic length, and the same was found for the two-band materials $2$H-NbSe$_{1.8}$S$_{0.2}$ and $2$H-NbS$_2$~\cite{Fente2016}. Theoretical results obtained in Ref.~\onlinecite{Ichioka2017} suggest that these materials can be in the so-called locking regime at which  $\xi_{\Delta 1}$ and  $\xi_{\Delta 2}$ are nearly the same. Microscopic calculations for clean two-band systems~\cite{Saraiva2017,Chen2020} demonstrate that this regime takes place when the interband (pair-exchange) coupling $g_{12}$ exceeds its locking value $g^*_{12}$. This value depends on the temperature $T$ and the ratio of the band Fermi velocities $v_{F2}/v_{F1}$. One finds~\cite{Saraiva2017,Chen2020} that $g_{12}^*$ drops significantly near $T_c$ and for the ratio $v_{F2}/v_{F1}$ close to $1$. Therefore, the two-band superconductors tend to be in the locking regime when approaching $T_c$ or in the case with nearly the same band Fermi velocities. Here it is worth noting that the ratio of the band Fermi velocities is indeed close to $1$~\cite{Tissen2013} in $2$H-NbS$_2$. 

Although researchers in the field of superconductivity are used to associate the spatial scale of a superconducting condensate with the corresponding gap function, the wider readership is more familiar with the Cooper-pair wave function. Therefore, a simple related question may arise whether or not it is more natural to use the characteristic length of the center-of-mass Cooper-pair wave function $\Psi_j({\bf r})$~[the Cooper-pair distribution] to determine the condensate length. In fact, this question is neither naive nor idle. In the case of conventional single-band superconductors, the gap function coincides with the center-of-mass wave function of a Cooper pair up to the Gor'kov coupling constant. It means that the gap-function and wave-function lengths are the same in this case. The situation is not that trivial for multiband superconductors. 

In the present work we investigate the gap-function $\xi_{\Delta j}$ and wave-function $\xi_{\Psi j}$ lengths for a two-band superconductor (with $j=1,2$), considered as a prototype of multiband superconductors. The lengths are calculated by numerically solving the $s$-wave microscopic formalism for an isolated vortex in the clean limit. Our study demonstrates that contrary to the single-band case, the difference between $\xi_{\Delta j}$ and $\xi_{\Psi j}$ is negligible only near the critical temperature $T_c$. At lower temperatures $\xi_{\Psi j}$ significantly deviates from $\xi_{\Delta j}$~(this deviation is not pronounced only for  $v_{F2}/v_{F1} \simeq 1$) and the fundamental question arises as to which of these lengths should be preferred when specifying the spatial scale of a band condensate in multiband superconducting materials.

\section{Formalism}
\label{sec:For}

\subsection{Gap function and Cooper-pair wave function}
\label{sec:For1}

When the pairing of electrons residing in different bands is negligible (which occurs in most cases), the number of the contributing condensates in a multiband superconductor is equal to the number of bands and the position dependent gap function $\Delta_i({\bf r})$~(for the $s$-wave pairing) is expressed in the form~\cite{Suhl1959,Moskalenko1959,Shanenko2011} 
\begin{eqnarray}
\Delta_i({\bf r}) = \sum_j g_{ij} \Psi_j({\bf r}),
\label{Delta}
\end{eqnarray}
where $g_{ij}$ is the coupling matrix and for the the center-of-mass wave function of a Cooper pair~\cite{Bogoliubov1970,Cherny1999} in band $j$ we have 
\begin{equation}
\Psi_j({\bf r})=\langle \hat{\psi}_{j\uparrow}({\bf r})\hat{\psi}_{j\downarrow}({\bf r}) \rangle,
\label{Phi}
\end{equation}
which is also referred to as the anomalous Green function of the field operators.  If $\Psi_j({\bf r})$ are proportional to the same position-dependent function $\psi({\bf r})$, one can immediately see from Eq.~(\ref{Delta}) that the band gap functions $\Delta_j({\bf r})$ are also exactly proportional to $\psi({\bf r})$. In this case, obviously, the gap-function spatial length $\xi_{\Delta i}$ is equal to the wave-function length $\xi_{\Psi i}$. This is the effectively single-condensate regime when the partial (band) condensates in a multiband system are characterized by the same spatial profile. However, when the contributing condensates are specified by different spatial distributions, one is not able to make any general statement about $\xi_{\Delta i}$ and $\xi_{\Psi i}$ without additional studies. 

\subsection{Bogoliubov-de Gennes equations and single vortex solution}
\label{sec:For2}

To clarify the issue about $\xi_{\Delta i}$ and $\xi_{\Psi i}$,  we consider the two-band generalization of the BCS model~\cite{Suhl1959,Moskalenko1959} with the $s$-wave pair condensates in both bands governed by the symmetric coupling matrix $g_{ij}\,(i,j=1,2)$. To solve the microscopic formalism, we employ the Bogoliubov-de Gennes (BdG) equations, which for the case of interest are written in the form~\cite{Komendova2012}
\begin{equation}\label{bdg}
\left[
\begin{array}{cc}
\hat{H}_{ei} & \Delta_i({\bf r}) \\
\Delta_i^*({\bf r}) & -\hat{H}^*_{ei}
\end{array}
\right] \left[
\begin{array}{c}
u_{i\nu}({\bf r}) \\
v_{i\nu}({\bf r})
\end{array}
\right] = E_{i\nu}
\left[
\begin{array}{c}
u_{i\nu}({\bf r}) \\
v_{i\nu}({\bf r})
\end{array}
\right],
\end{equation}
where $u_{i\nu}(\bf r)$ and $v_{i\nu}({\bf r})$ are the electron-like and hole-like wave functions associated with band $i$ ($\nu$ is the set of the relevant quantum numbers), $E_{i\nu}$ represents the quasiparticle energies, and $\hat{H}_{ei}$ is the single-particle Hamiltonian (absorbing the chemical potential). For our calculations we assume the effective mass approximation and quasi-2D bands, as emergent multiband superconductors often exhibit quasi-2D Fermi surfaces~\cite{Paglione2010}, so that 
$
\hat{H}_{ei}({\bf r})= -\frac{\hbar^2}{2m_i}\big(\partial^2_x + \partial^2_y\big)  - \mu_i,
\label{T}
$
where the single-particle energy is degenerate in the $z$ direction, $m_i$ is the electron band mass, $\mu_i=m_iv_{Fi}^2/2$ is the chemical potential measured from the lower edge of the corresponding band, with $v_{Fi}$ the band Fermi velocity. The magnetic field is not included as we consider the system in the deep type II regime, where the magnetic field does not vary within the spatial scales of the contributing condensates and cannot influence their spatial lengths. 
In addition, for the sake of simplicity we ignore impurities, considering the clean limit. 

The BdG equations are solved in the self-consistent manner, together with Eq.~(\ref{Delta}) and the relation
\begin{equation}
 \Psi_j({\bf r}) = \sum_{\nu} u_{j\nu}({\bf r}) v^*_{j\nu}({\bf r})\big[1-2f(E_{j\nu})\big],
\label{self}
\end{equation}
where $f(E_{j\nu})$ is the Fermi-Dirac distribution of bogolons and the summation includes the states with positive quasiparticle energies for which the single-electron energy falls in the Debye window $[\mu_j-\hbar \omega_D,~\mu_j+\hbar\omega_D]$, with $\omega_D$ the Debye frequency.  As is seen, the pair-exchange coupling between the two contributing bands $g_{12}=g_{21}$ is not explicitly present in the BdG equations but appears in the self-consistency equation (\ref{Delta}). We remark that to go beyond the adopted model, one should take into account Cooper pairs made of electrons from different bands, see e.g. Ref.~\onlinecite{Shanenko2015, Vargas2020}. In this case the coupling between bands  $1$ and $2$ appears explicitly in the BdG equations~\cite{Shanenko2015,Vargas2020}. However, as is mentioned above, in most cases the interband pairing is negligible and we do not consider this point in our present work. 

To find $\xi_{\Psi j}$ and $\xi_{\Delta j}$, we adopt an isolated vortex oriented along the $z$ direction~\cite{Gygi1991, Hayashi1998} with
\begin{equation}
\Delta_j({\bf r}) =\Delta_j(\rho)e^{-i\theta},\;\Psi_j({\bf r})=\Psi_j(\rho)e^{-i\theta},
\label{vortex}
\end{equation}
where $\rho,\theta,z$ are the cylindrical coordinates, and $\Delta_j(\rho)$ and $\Psi_j(\rho)$ are the radial parts of the band gap functions and Cooper-pair center-of-mass wave functions, respectively. These expressions are inserted into the BdG equations and then, the equations are converted into the matrix form by means of the expansion in a set of the single-electron wave functions. Further details of the numerical procedure can be found in the Supplemental Material and also in Ref.~\onlinecite{Chen2020}. 

In our calculations for band $1$ we choose $\mu_1 = 30$ meV, which is in the range of the Fermi energies in recent multiband superconductors~\cite{Lubashevsky2012}. The chemical potential relative to the lower edge of band $2$~($\mu_2$) is varied in order to consider different values of the ratio $v_{F2}/v_{F1}$. We also use the cut-off energy $\hbar\omega_D = 15$ meV and the dimensionless intraband coupling $g_{11}N_1 = 0.3$, with $N_i$ the density of states (these values are typical of the conventional superconductors~\cite{Fetter}). For band $2$ we take $g_{22}=0.8g_{11}$. The two values $g_{12}=0.3g_{11}$ and $0.9g_{11}$ are considered for the pair-exchange (interband) coupling to check both the weak- and strong-coupling regimes of the intraband interactions. We note that most of multiband superconductors exhibit weak pair-exchange couplings but there are also examples with relatively strong intraband interactions like MgB$_{2}$, see e.g. table II of Ref.~\onlinecite{Vagov2016} and also Refs.~\onlinecite{Golubov2002,Singh2010,Khasanov2010,Kim2011}.

\subsection{Healing lengths}
\label{sec:For3}

It is also of importance to discuss how the gap- and wave-function lengths can be extracted from the numerical solution of the BdG equations for an isolated vortex. There are different practical definitions of $\xi_{\Delta}$ utilized in the single-band case. It can be defined as the size of the vortex core related to the gap function slope~\cite{Bardeen1969, Caroli1964,Schmid1966, Clem1975}, or the radius of the maximal supercurrent density~\cite{Sonier2004}, or the radius of a cylinder containing the energy equal to the condensation energy~\cite{Gennes1966, Tinkham1996}. One can also identify $\xi_{\Delta}$ as the healing length, i.e. the distance over which the gap function can nearly heal, being suppressed in the center of a vortex. In the latter case $\xi_{\Delta}$ can be found from the fact that the radial profile of the gap function is well approximated~\cite{Schmid1966, Clem1975} as $\Delta(\rho)=\Delta_0\rho/\sqrt{\rho^2 + \xi^2_{\Delta}}$, with $\Delta_0$ the bulk value of the gap function.  

In our study we employ the healing-length approach, and obtain the characteristic lengths $\xi_{\Delta j}$ and $\xi_{\Psi j}$ by fitting to the numerical results with the approximations
\begin{equation}
\Delta_j(\rho) \simeq \frac{\Delta_{j0}\rho}{\sqrt{\rho^2 + \xi^2_{\Delta j}}}, \; \Psi_j(\rho) \simeq \frac{\Psi_{j0}\rho}{\sqrt{\rho^2 + \xi^2_{\Psi j}}},
\label{app}
\end{equation}
where $\Delta_{j0}$ and $\Psi_{j0}$ are the values of $\Delta_j(\rho)$ and $\Psi_j(\rho)$ far beyond the vortex core. We stress that the approximations of Eq.~(\ref{app}) are not used to solve the BdG equations. Equation (\ref{app}) is employed to get the characteristic lengths from the exact numerical solution of the BdG equations. Notice that nearly the same results for the characteristic lengths can be obtained from the criterion
\begin{equation}
\Delta_j(\rho=\xi_{\Delta j})=\frac{\Delta_{j0}}{\sqrt{2}}, \; \Psi_j(\rho=\xi_{\Psi j})=\frac{\Psi_{j0}}{\sqrt{2}},
\label{app1}
\end{equation}
where the choice of the factor $\sqrt{2}$ is justified by Eq.~(\ref{app}). We also remark that using the healing lengths is not crucial for our results, our conclusions are general and not sensitive to particular definitions of $\xi_{\Delta j}$ and $\xi_{\Psi j}$, see the Supplemental Material.   

\section{Results}
\label{sec:Res}

\begin{figure}
	\begin{center}
		\includegraphics[width=1.0\linewidth]{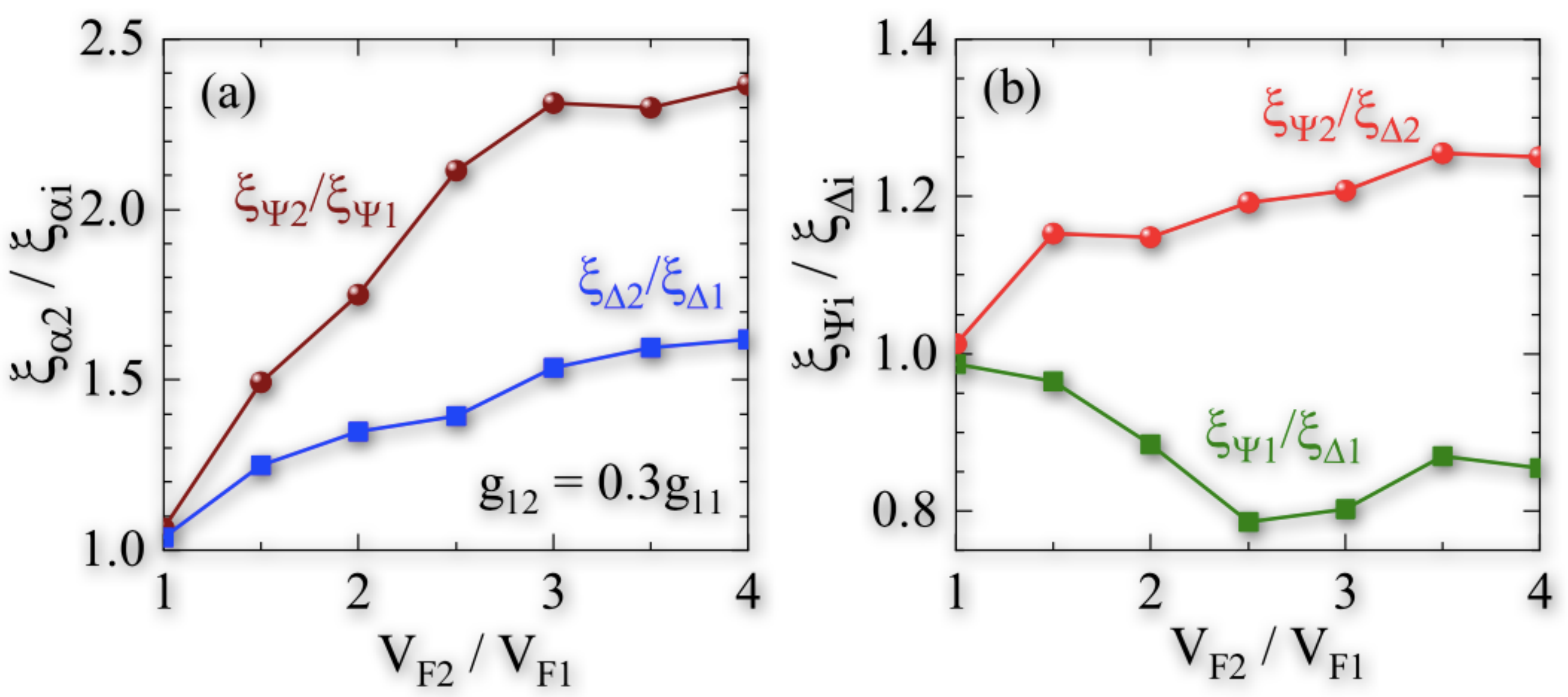}
	\end{center}
	\caption{a) The ratio $\xi_{\alpha 2}/\xi_{\alpha 1}$~(for $\alpha=\Delta,\Psi$) versus the ratio of the band Fermi velocities $v_{F2}/v_{F1}$ for weakly coupled condensates $g_{12}=0.3g_{11}$. b) $\xi_{\Psi i}/\xi_{\Delta i}$~(for $i=1,2$) as a function of $v_{F2}/v_{F1}$. The lines simply connect the dots and serve as a guide for the eyes.}
	\label{fig1}
\end{figure}

\subsection{Numerical results of the BdG equations}
\label{sec:Res1}

Let us first consider results for weakly coupled condensates with $g_{12}=0.3g_{11}$. Figure \ref{fig1} (a) demonstrates the ratios $\xi_{\Delta 2}/\xi_{\Delta 1}$ and  $\xi_{\Psi 2}/\xi_{\Psi 1}$ as functions of $v_{F2}/v_{F1}$ at $T=0$. One can see that when $v_{F2}/v_{F1}$ is close to $1$, the system approaches the locking regime so that  $\xi_{\Delta 2}/\xi_{\Delta 1}\approx \xi_{\Psi 2}/\xi_{\Psi 1}\approx 1$. When $v_{F2}/v_{F1}$ increases, we obtain different condensate lengths. However, the ratio $\xi_{\Delta 2}/\xi_{\Delta 1}$ becomes notably smaller than $\xi_{\Psi 2}/\xi_{\Psi 1}$. For example, at $v_{F2}/v_{F1}=4$ we have $\xi_{\Delta 2}/\xi_{\Delta 1}=1.6$ while $\xi_{\Psi 2}/\xi_{\Psi 1}=2.4$. 

Further insight is provided by Fig.~\ref{fig1}(b), where $\xi_{\Psi 2}/\xi_{\Delta 2}$ and $\xi_{\Psi 1}/\xi_{\Delta 1}$ are given versus  $v_{F2}/v_{F1}$. As is seen, for  $v_{F2}/v_{F1}=1$ all the four lengths are nearly the same, in agreement with the data given in Fig.~\ref{fig1}(a). When increasing $v_{F2}/v_{F1}$, the wave-function lengths $\xi_{\Psi i}$ systematically deviate from the gap-function lengths $\xi_{\Delta i}$. Moreover, the condensates in bands $1$ and $2$ show opposite trends. For band $2$ we have $\xi_{\Psi 2} > \xi_{\Delta 2}$ and the ratio $\xi_{\Psi 2}/\xi_{\Delta 2}$ exhibits an overall growth when increasing $v_{F2}/v_{F1}$ in Fig.~\ref{fig1}(b). On the contrary, for band $1$ we have $\xi_{\Psi 1} < \xi_{\Delta 1}$ and the ratio $\xi_{\Psi 1}/\xi_{\Delta 1}$ decreases with increasing $v_{F2}/v_{F1}$. One sees that at $v_{F2}/v_{F1}=4$ both $\xi_{\Psi 1}$ and  $\xi_{\Psi 2}$ deviate from the corresponding gap-function lengths by about $20-25\%$. Thus, we arrive at the striking conclusion that each condensate in the model of interest is specified by two generally different lengths $\xi_{\Psi i}$ and $\xi_{\Delta i}$.

\begin{figure}
\centering
\includegraphics[width=1.0\linewidth]{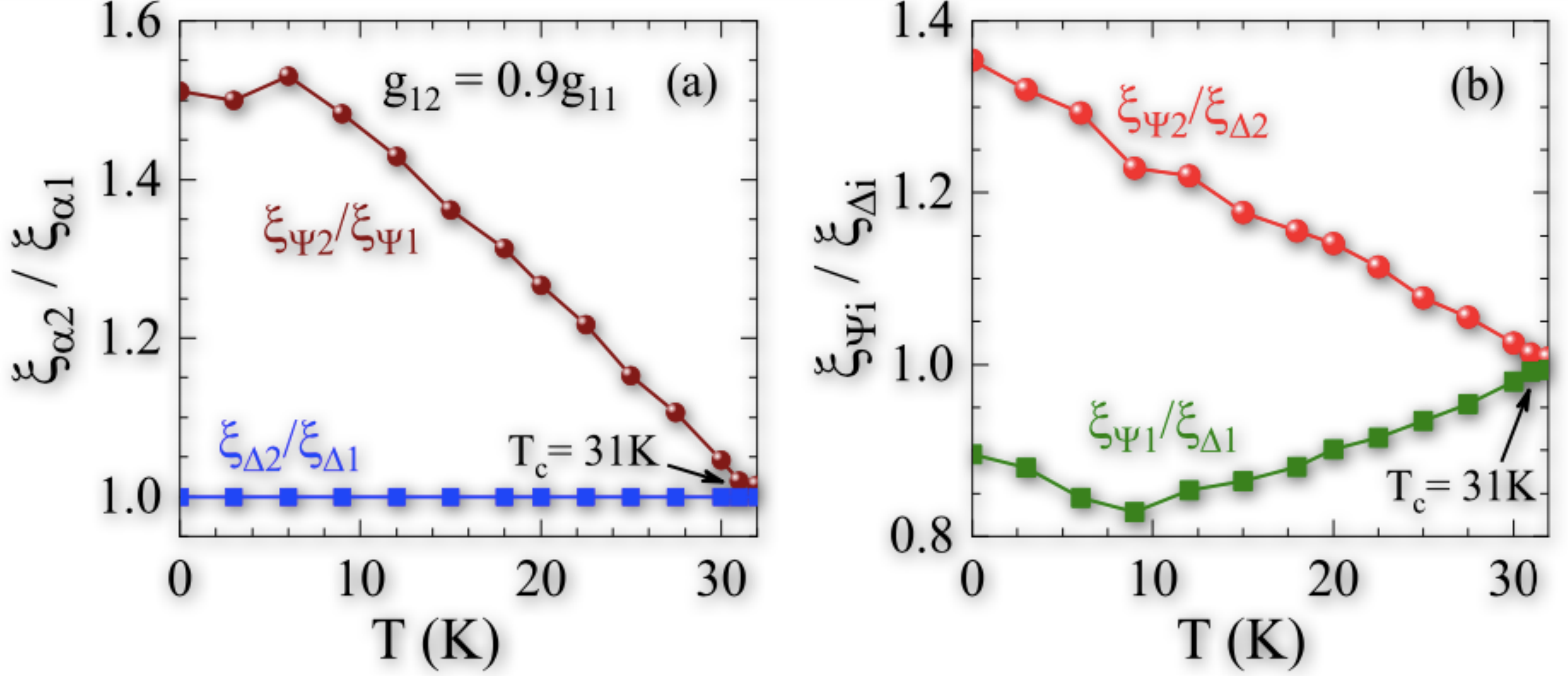}
\caption{The same as in Fig.~\ref{fig1} but versus the temperature and for strongly coupled condensates with $g_{12}=0.9g_{11}$ and $v_{F2}/v_{F1}=\sqrt{3}$. Lines simply join points and are a guide to the eyes.
}\label{fig2}
\end{figure}

Our results demonstrated in Fig.~\ref{fig1} are obtained for weakly coupled bands. Now we turn to the case of strongly coupled condensates. Based on the previous studies~\cite{Ichioka2017,Saraiva2017,Chen2020}, we know that the difference between the gap-function lengths $\xi_{\Delta 1}$ and $\xi_{\Delta 2}$ decreases with increasing $g_{12}$ (at a constant ratio $v_{F2}/v_{F1}$). Then, one expects that in the case of strongly coupled bands the gap-function lengths should be close to one another, as the system approaches the locking regime. Here the question arises as for whether or not the wave-function lengths $\xi_{\Psi 1}$ and $\xi_{\Psi 2}$ exhibit the same trend. To answer this question, Fig.~\ref{fig2} demonstrates the data calculated from the BdG equations for different temperatures at $g_{12}=0.9g_{11}$ and for the same intraband coupling constants $g_{11}$ and $g_{22}$ as in Fig.~\ref{fig1}. Here we set $\mu_2/\mu_1=3$, which corresponds to $v_{F2}/v_{F1}=\sqrt{3}$. This moderate ratio of the band Fermi velocities ensures that the characteristic lengths of the band-dependent gap functions are nearly in the locking regime.

The temperature dependent quantities $\xi_{\Psi 2}/\xi_{\Psi 1}$ and $\xi_{\Delta 2}/\xi_{\Delta 1}$ are shown in Fig.~\ref{fig2}(a). One sees that the gap-function lengths of bands $1$ and $2$ are indeed almost the same. However, $\xi_{\Psi 2}$ and $\xi_{\Psi 1}$ differ notably from one another at relatively low temperatures. The ratio $\xi_{\Psi 2}/\xi_{\Psi 1}$ approaches $1$ only in the vicinity of the critical temperature $T_c=31 {\rm K}$. At first sight the results for the wave-function lengths given in Fig.~\ref{fig2}(a) looks weird because, as is mentioned in the discussion after Eq.~(\ref{Delta}), the equality  $\xi_{\Delta 1}=\xi_{\Delta 2}$ assumes the equal wave-function lengths $\xi_{\Psi 1}=\xi_{\Psi 2}$. In fact, the key point here is that in our case the gap-function lengths $\xi_{\Delta 1}$ and $\xi_{\Delta 2}$ are slightly different: at low temperatures ($< 10\,{\rm K}$) this difference is about one percent~(see the spatial profiles of $\Delta_j(\rho)$ in the Supplemental Material). Strikingly, this almost negligible deviation between the two gap-function lengths results in an order-of-magnitude larger difference between the wave-function lengths $\xi_{\Psi 1}$ and $\xi_{\Psi 2}$. 

To have a feeling about this result, let us consider Eq.~(\ref{Delta}) for the simplified case $g_{11}=g_{22}$~(close to our choice). Then, one can find from Eq.~(\ref{Delta}) that $\Delta_2({\bf r}) -\Delta_1({\bf r})= (1-g_{12}/g_{11}) [g_{11}\Psi_2({\bf r}) - g_{11}\Psi_1({\bf r})]$. As $1-g_{12}/g_{11}=0.1$, one can conclude that the difference between the band gaps is an order of magnitude smaller than the difference between $g_{11}\Psi_1$ and $g_{11}\Psi_2$. At the same time $\Delta_2$, $\Delta_1$, $g_{11}\Phi_1$, and $g_{11}\Phi_2$ are of the same order of magnitude. It means that for similar $\xi_{\Delta 1}$ and $\xi_{\Delta 2}$, the characteristic lengths of the spatial variations of $g_{11} \Psi_1({\bf r})$ and $g_{11}\Psi_2({\bf r})$ are very different. Finally, it only remains to take into account that $g_{11}\Psi_1({\bf r})$ and $g_{11}\Psi_2({\bf r})$ have the same spatial lengths as $\Psi_1({\bf r})$ and $\Psi_2({\bf r})$, respectively.

As $\xi_{\Psi 2}/\xi_{\Psi 1}$ differs significantly from $\xi_{\Delta 2}/\xi_{\Delta 1}$, we must have notably different  $\xi_{\Psi i}$ and $\xi_{\Delta i}$. For more detailed information, Fig.~\ref{fig2}(b) demonstrates $\xi_{\Psi 1}/\xi_{\Delta 1}$ and $\xi_{\Psi 2}/\xi_{\Delta 2}$ as functions of the temperature, calculated for the same microscopic parameters as the results given in Fig.~\ref{fig2}(a). One sees that near $T_c$ both quantities are close to $1$. However, when the temperature decreases, we find that $\xi_{\Psi 2}$ deviates significantly upward from $\xi_{\Delta 2}$ while $\xi_{\Psi 1}$ becomes smaller then $\xi_{\Delta 1}$. The results in Fig.~\ref{fig2} are obtained for $v_{F2}/v_{F1}=\sqrt{3}$. For a larger difference between $v_{F1}$ and $v_{F2}$, we obtain even larger deviations between the gap- and wave-function lengths of the same band. When $v_{F2}$ is close to $v_{F1}$, the system approaches the length locking regime for the both gap- and wave-function lengths, in agreement with the results for the weakly-coupled bands in Fig.~\ref{fig1}.  

\subsection{Local relation between the lengths}
\label{sec:res2}

Our study would not be complete without discussing a simplified analytical relation between the gap- and wave-function lengths which can be obtained from the approximations given by Eq.~(\ref{app}). When using Eq.~(\ref{app}), one finds for $\rho \to 0$
\begin{align}
\Delta_j(\rho)  = \rho \frac{\Delta_{j0}}{\xi_{\Delta j}}, \; \Psi_j(\rho) = \rho \frac{\Psi_{j0}}{\xi_{\Psi j}}.
\label{app2}
\end{align}
Inserting Eq.~(\ref{app2}) into Eq.~(\ref{Delta}), one gets
\begin{align}
\frac{\Delta_{i0}}{\xi_{\Delta i}} =\sum\limits_j g_{ij}\frac{\Psi_{j0}}{\xi_{\Psi j}}.  
\label{Delta1}
\end{align} 
Solving these equation for the two-band case together with Eq.~(\ref{Delta}) taken for $\rho \to \infty$ , one gets  
\begin{align}
\xi_{\Psi 1}=\frac{\xi_{\Delta 1} \xi_{\Delta 2}\big(\gamma_{11}\Delta_{10} + \gamma_{12}\Delta_{20}\big)}{\gamma_{11}\Delta_{10}\xi_{\Delta 2}+\gamma_{12}\Delta_{20}\xi_{\Delta 1}},\notag\\ 
\xi_{\Psi 2}=\frac{\xi_{\Delta 1} \xi_{\Delta 2}\big(\gamma_{21}\Delta_{10} + \gamma_{22}\Delta_{20}\big)}{\gamma_{21}\Delta_{10}\xi_{\Delta 2}+\gamma_{22}\Delta_{20}\xi_{\Delta 1}},
\label{xiDelPsi}
\end{align} 
where $\gamma_{ij}$ are elements of the inverse coupling matrix. We stress that Eq.~(\ref{xiDelPsi}) is an approximation. Indeed, one can utilize Eq.~(\ref{Delta}) at any value of $\rho$ to express $\xi_{\Psi i}$ in terms of $\xi_{\Delta i}$ by using Eq.~(\ref{app}). Since the latter is approximative, new expressions for  $\xi_{\Psi 1}$ and $\xi_{\Psi 2}$ will be different from Eq.~(\ref{xiDelPsi}). Nevertheless, the difference will not be significant because the fitting to the numerical results for the gap- and wave functions demonstrates that the approximations of Eq.~(\ref{app}) are quite good~(see the Supplemental Material). This makes it possible to find the wave-function lengths from the gap-function lengths, if one, of course, knows the coupling matrix and bulk gaps $\Delta_{j0}$. Once the gap-function lengths have been extracted from the STM measurements, use of Eq.~(\ref{xiDelPsi}) opens the way to the derivation of the wave-function lengths. However, the accuracy of estimating the gap-function lengths from the experimental results should be high enough to get reliable values of the wave-function lengths, see the discussion of our results in Fig.~\ref{fig2}.

\section{Conclusions}

Concluding, we have investigated the characteristic lengths of two coupled condensates in a two-band superconducting material with the $s$-wave pairing in both contributing bands. Our results for the lengths have been extracted from the exact numerical solution of the two-band BdG equations for a single vortex by using the healing-length approach. For each contributing condensate one of the lengths is related to the center-of-mass Cooper pair wave function while another is associated with the corresponding gap function. Though there is no difference between such lengths in the single-band case, our study has demonstrated that for two-band superconductors this is true only in the vicinity of the critical temperature. At lower temperatures the gap- and wave-function lengths systematically deviate from each other, and this deviation is more pronounced with increasing the difference between the band Fermi velocities. In this respect we remark that the ratio of the band Fermi velocities can differ significantly from $1$ in two-band superconductors, for example, $v_{F2}/v_{F1}$ is about~\cite{Suderow2005} $0.05$ in 2H-NbS$_2$ whereas it is close~\cite{Tissen2013} to $18.2$ in 2H-NbSe$_2$.

As is mentioned above, multiple condensates in one system can interfere constructively or destructively, which results in unconventional coherent behavior.  Since the interference effects in a multiband system are controlled by the Cooper-pair wave function of the aggregate condensate, one can expect that the wave-function lengths should be preferred when considering the relevant spatial scales of the contributing condensates. However, additional studies, e.g., including possible analytical results within the exact perturbative analysis of the microscopic equations in the vicinity of $T_c$, are certainly necessary to further clarify this fundamental issue of multiband superconductors. 

\section*{Acknowledgements}
This work was supported by Zhejiang Provincial Natural Science Foundation (Grant No. LY18A040002) and Science Foundation of Zhejiang Sci-Tech University(ZSTU) (Grant No. 19062463-Y). The work at HSE University (A.A.S.) was financed within the framework of the Basic Research Program of HSE University.

\appendix*
\section{SUPPLEMENTAL MATERIAL}

\subsection{Self-consistent solution of the Bogoliubov-de Gennes equations}

Our consideration is based on the previous studies of an isolated vortex within the single-band~\cite{Bardeen1969, Gygi1991,Hayashi1998} and two-band BdG equations~\cite{Komendova2012, Araujo2009}. Following these papers and considering a single vortex oriented along the $z$ direction, we represent the particle- and hole-like wave functions in the form 
\begin{eqnarray}
u_{i,\nu }({\bf r}) &=& \frac{1}{\sqrt{2\pi
L}}u_{i,jm}(\rho)e^{\mathbbm{i}(m-\frac{1}{2})\theta}e^{\mathbbm{i}k_zz}, \nonumber \\
v_{i,\nu}({\bf r}) &=& \frac{1}{\sqrt{2\pi
L}}v_{i,jm}(\rho)e^{\mathbbm{i}(m+\frac{1}{2})\theta}e^{\mathbbm{i}k_zz},\label{uv}
\end{eqnarray}
where $\rho,\theta$ and $z$ are the cylindrical coordinates, $L$ is the unit cell of the periodic boundary conditions in the $z$-direction, and $\nu=\{j, m, k_z\}$, with $j$ the radial quantum number, $m$ the azimuthal quantum number being half an odd integer, and $k_z$ the wavenumber in the $z$-direction. For our calculations we assume the effective mass approximation and quasi-2D bands with the single-particle energies degenerate in the $z$ direction (see the article). In this case the radial parts of the wave functions $u_{i,\nu }(\rho)$ and $v_{i,\nu }(\rho)$ do not depend on $k_z$. 

The BdG equations are solved together with the boundary conditions $u_{i,jm}(\rho=R)=0$ and $v_{i,jm}(\rho =R) = 0$, i.e., a single vortex is considered in a cylinder with the axial direction along the $z$ axis. To eliminate the quantum confinement effects we choose the sufficiently large radius $R=300\,{\rm nm}$ that is orders of magnitude larger than the Fermi wavelengths of the both contributing bands. 

To represent the BdG equations in the matrix form, we expand the radial parts of the radial wave functions $u_{i,jm}(\rho)$ and $v_{i,jm}(\rho)$ in terms of the normalized Bessel functions of the first kind
\begin{equation}
\phi_{jm}^{(\pm)}(\rho)= \frac{\sqrt{2}}{R\mathcal{J}_{(m+1)\pm\frac{1}{2}}(\alpha_{j,m\pm\frac{1}{2}})}
\mathcal{J}_{ m\pm\frac{1}{2}} \big(\alpha_{j,m\pm\frac{1}{2}}\frac{\rho}{R}\big),
\label{phi}
\end{equation}
where ``-" and ``+" in the superscripts are for $u$ and $v$ functions, respectively, and $\alpha_{j,\eta}$ is given by $J_{\eta}(\alpha_{j,\eta})=0$. The expansion writes
\begin{align}
u_{i,jm}(\rho) &=\sum\limits_{j'=1}^N c_{i,jj'm}\phi^{(-)}_{j'm}(\rho),\notag\\
v_{i,jm}(\rho) &=\sum\limits_{j'=1}^N d_{i,jj'm}\phi^{(+)}_{j'm}(\rho) ,
\label{exp}
\end{align}
where the number of the Bessel functions $N$ is large enough; in our calculations we employed the values $N > 100$. As a result, we get the BdG equations (\ref{bdg}) in the form of the matrix ($2N\times 2N$) equation with the elements of the eigenvectors given by $c_{i,jj'm}$ in the upper half of the column and by $d_{i,jj'm}$ in the lower half of the column. Then, the problem is solved in the self consistent manner. At the first step we choose (as a scientific guess) initial gap functions $\Delta_i(\rho)$, calculate the corresponding $2N\times 2N$-matrix and find the eigenstates and eigenvalues of the matrix BdG equation. Second, we utilize the obtained quantities $c_{i,jj'm},\,d_{i,jj'm}$, and $E_{i,jm}$ to find new gap functions by using Eqs.~(\ref{Delta}), (\ref{Phi}) and (\ref{self}) from the article. Third, we solve the BdG equations with the new gap functions, to derive a new set of $c_{i,jj'm}$, $d_{i,jj'm}$, and $E_{i,jm}$. The calculations are repeated until the convergence is reached with the numerical residual $10^{-6}$.

\subsection{Microscopic parameters}

As is mentioned in the article, the couplings constants $g_{ij}$ are chosen in units of $g_{11}$, while $g_{11}N_1 = 0.3$, with $N_i$ the band-dependent DOS. Here we explain how the degenerate states in the $z$ direction are taken into account when calculating $N_1$ and also performing the summation in Eq.~(\ref{self}) of the article. We have $N_i=(m_1/2\pi\hbar^2 L) \sum_{k_z}\theta(k_{\rm max} - |k_z|)$, with $\theta(k_{\rm max} - |k_z|)$ the step function and $k_{\rm max}$ the maximal wavenumber in the $z$ direction. [The effective band-dependent electron masses $m_i$ are set to the free electron mass $m_e$, for simplicity.] One can employ the estimate $k_{\rm max} = \pi/a_z$, where $a_z$ is the lattice constant. Then, we get $(1/L) \sum_{k_z}\theta(k_{\rm max} - k_z) \approx 1/a_z$. For typical values for the lattice constant $1/a_z \sim 1$-$3 {\rm nm}^{-1}$. For our calculations choose $N_i= \tilde{N} m_e/ 2\pi\hbar^2$, with $\tilde{N}=1 {\rm nm}^{-1}$. This way a particular value of $L$ is hidden in $\tilde{N}$. The same quantity appears when the summation over $k_z$ is performed in Eq.~(\ref{self}) of the article. Notice that the choice of $\tilde{N}$ and also the use of $m_i=m_e$ do not influence our conclusions  because any changes in $N_i$ result simply in the trivial renormalization of the couplings $g_{11}$, $g_{22}$, and $g_{12}$. 

\begin{figure}
	\begin{center}
		\includegraphics[width=1.0\linewidth]{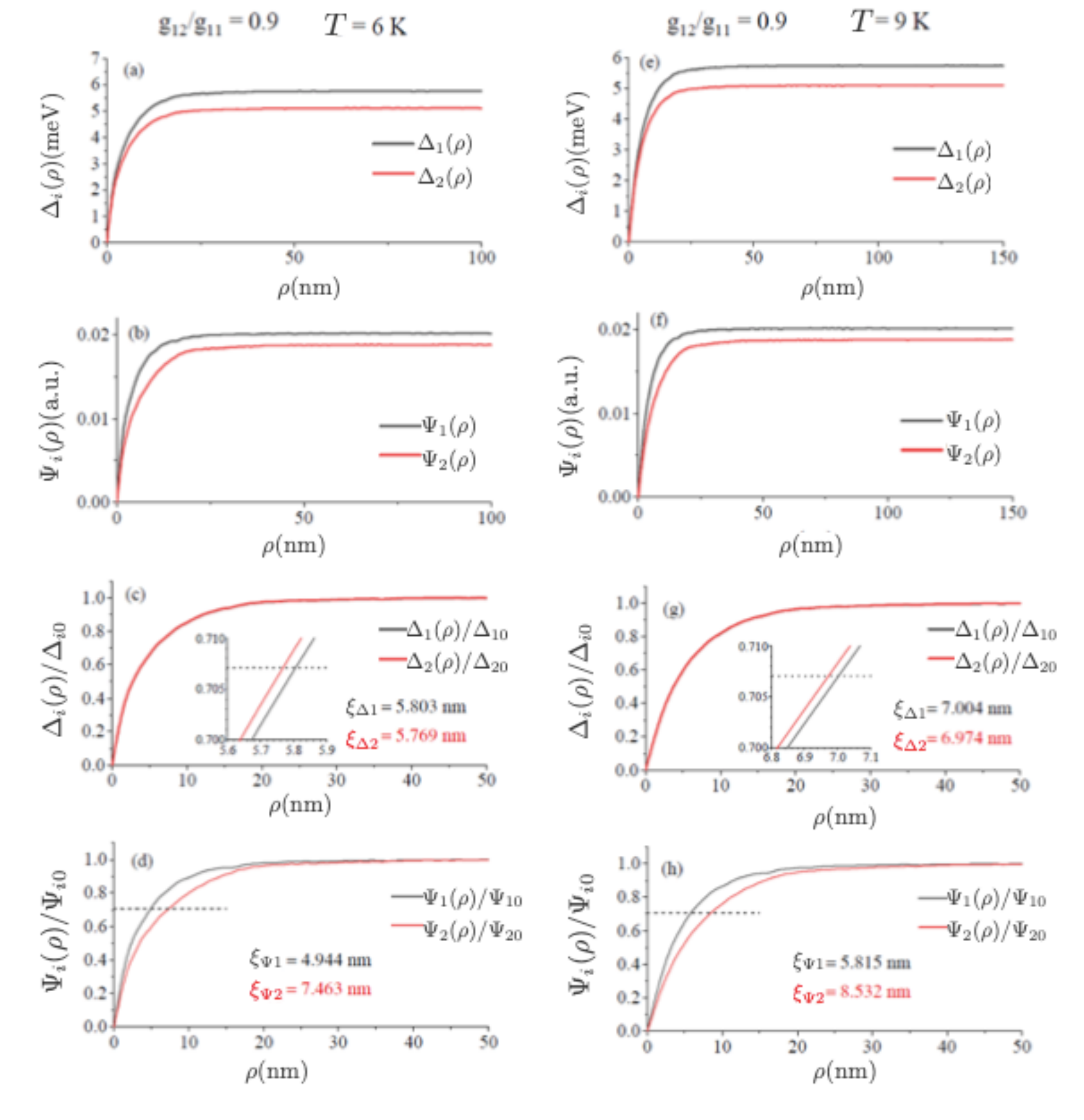}
	\end{center}
	\caption{The radial-dependent parts of the gap and wave functions of the both contributing condensates [i. e. $\Delta_i(\rho)$ and $\Psi_i(\rho)$] for $T=6\, {\rm K}$ [panels (a)-(d)] and $9\,{\rm K}$ [panels (e)-(h)]; the calculations are performed for the strongly coupled bands with $g_{12}=0.9g_{11}$. Panels (c)-(d) and (g)-(h) illustrate how the gap- and wave-function lengths [i. e. $\xi_{\Delta i}$ and $\xi_{\Psi i}$] are extracted from $\Psi_i(\rho)/\Psi_{i0}$ and $\Delta_i(\rho)/\Delta_{i0}$~(the dotted lines mark the corresponding characteristic lengths). The inserts in panels (c) and (g) demonstrate an almost negligible difference between $\xi_{\Delta 1}$ and $\xi_{\Delta 2}$.}
	\label{fig3}
\end{figure}

\subsection{Examples of the gap-function and wave-function spatial distributions}

The radial parts of the gap and wave functions $\Delta_i(\rho)$ and $\Psi_i(\rho)$ are shown in Fig.~\ref{fig3}.
The horizontal dotted lines in panels (c), (d), (g), (h) serve as guides for the eyes to mark the characteristic lengths. In particular, one can see that the lengths derived by fitting to the numerical results with the approximations given by Eq.~(\ref{app}) in the article are very close to those obtained from the simplified conditions given by Eq.~(\ref{app1}). Figure~\ref{fig3} demonstrates that the wave-function lengths $\xi_{\Psi 1}$ and $\xi_{\Psi 2}$ are significantly different while $\xi_{\Delta 1}$ and $\xi_{\Delta 2}$ are nearly the same. For example, from Fig.~\ref{fig3}(c) we learn that $\xi_{\Delta 1}=5.803\,{\rm nm}$ and $\xi_{\Delta 2}=5.769\,{\rm nm}$ while from Fig.~\ref{fig3}(d) one finds $\xi_{\Psi 1}=4.944\,{\rm nm}$ and $\xi_{\Psi 2}=7.463\,{\rm nm}$. As is seen, the difference between the gap-function lengths $|\xi_{\Delta 1}-\xi_{\Delta 2}|/\xi_{\Delta 1} = 0.005$ while  $|\xi_{\Psi 1}-\xi_{\Psi 2}|/\xi_{\Psi 1} = 0.33$, in agreement with the discussion about the results given in Fig.~\ref{fig2}.  

One can also see from Figs.~\ref{fig3}(d) and (h) that the spatial profiles of the wave functions $\Psi_1(\rho)$ and $\Psi_2(\rho)$ are different while the spatial profiles of the gap-functions in Fig.~\ref{fig3}(c) and (g) are nearly the same. This difference demonstrates that our conclusions cannot be sensitive to a particular definition of the healing length. 

\begin{figure}
	\begin{center}
		\includegraphics[width=0.6\linewidth]{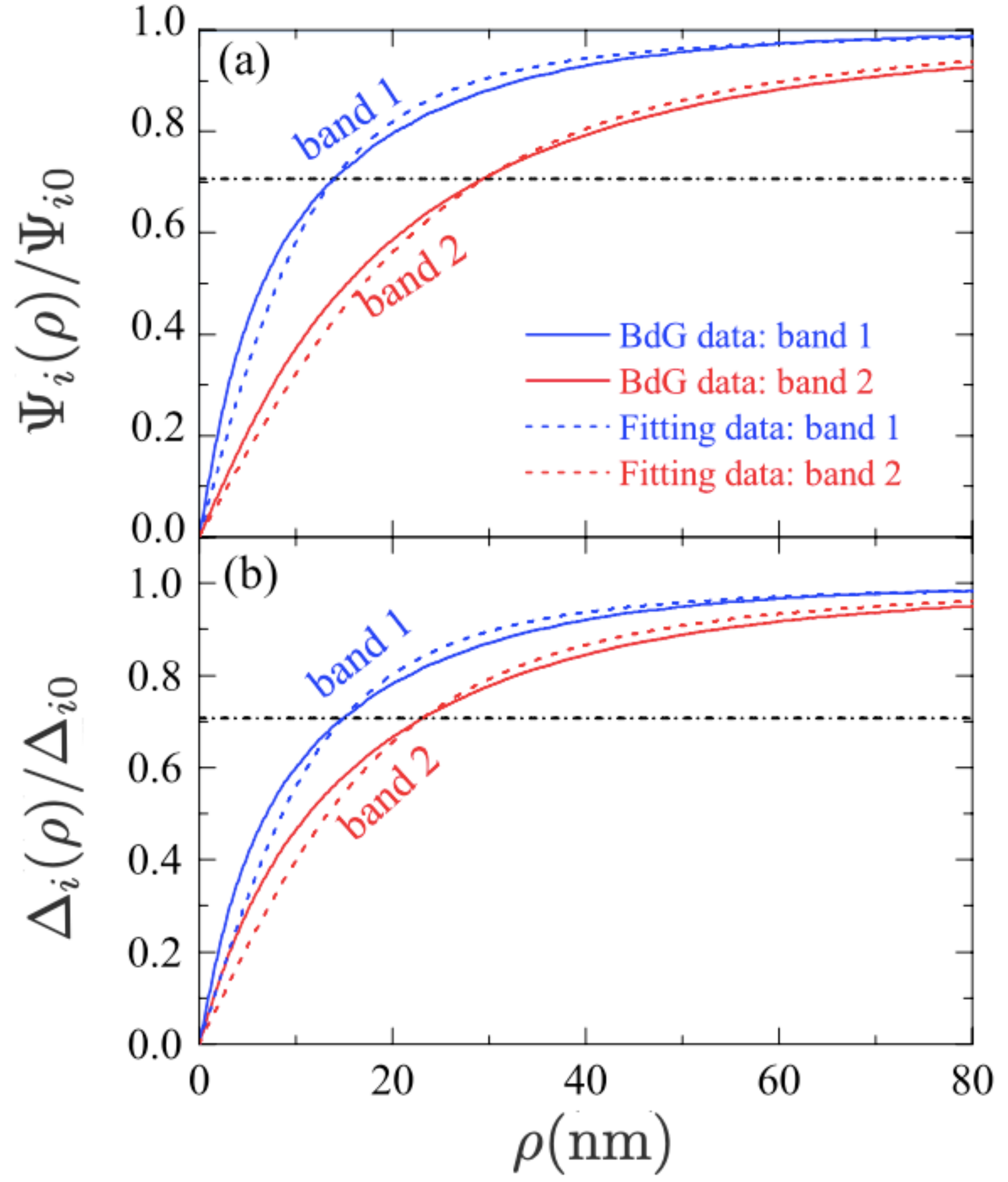}
	\end{center}
	\caption{The radial-dependent parts of the gap and wave functions of the both contributing condensates:
	the solid lines represent our numerical results and the dotted lines are the fitting results obtained by
	applying Eq.~(\ref{app}) in the article.}
	\label{fig4}
\end{figure}

In addition, Fig.~\ref{fig4} demonstrates that the approximations given by Eq.~(\ref{app}) are in good agreement with the gap- and wave functions in the core of the single vortex calculated within the BdG equations. 

\begin{thebibliography}{100}
\bibitem{MP2015} M. V. Milo\v{s}evi\'{c} and A. Perali, Emergent phenomena in multicomponent superconductivity: an introduction to the focus issue, Supercond. Sci. Technol. {\bf 28}, 060201 (2015).
\bibitem{Lin2014} S.-Z. Lin, Ground state, collective mode, phase soliton and vortex in multiband superconductors, Supercond. Sci. Technol. {\bf 26}, 493202 (2014).
\bibitem{Tanaka2015} Y. Tanaka, Multicomponent superconductivity based on multiband superconductors, Supercond. Sci. Technol. {\bf 28}, 034002 (2015).
\bibitem{Lubashevsky2012} Y. Lubashevsky, E. Lahoud, K. Chashka, D. Podolsky, and A. Kanigel, Shallow pockets and very strong coupling superconductivity in FeSe$_x$Te$_{1-x}$, Nat. Phys. {\bf 8}, 309 (2012).
\bibitem{Okazaki2014} K. Okazaki, Y. Ito, Y. Ota, Y. Kotani, T. Shimojima, T. Kiss, S. Watanabe, C.-T. Chen,S. Niitaka, T. Hanaguri, H. Takagi, A. Chainani, and S. Shin, Superconductivity in an electron band just above the Fermi level: possible route to BCS-BEC superconductivity, Sci. Rep. {\bf 4}, 4109 (2014).
\bibitem{Chen2012} Y. Chen, A. A. Shanenko, A. Perali and F. M. Peeters, Superconducting nanofilms: molecule-like pairing induced by quantum confinement, J. Phys.: Condens. Matter {\bf 24}, 185701 (2012).
\bibitem{Babaev2002} E. Babaev, Vortices with fractional flux in two-gap superconductors and in extended Faddeev model, Phys. Rev. Lett. {\bf 89}, 067001 (2002).
\bibitem{Bluhm2006} H. Bluhm, N. C. Koshnick, M. E. Huber, and K. A. Moler, Magnetic response of mesoscopic superconducting rings with two order parameters, Phys. Rev. Lett. {\bf 97}, 237002 (2006).
\bibitem{Lin2013} S. Z. Lin and C. Reichhardt, Stabilizing fractional vortices in multiband superconductors with periodic pinning arrays, Phys. Rev. B 87, 100508 (2013).
\bibitem{Tanaka2001} Y. Tanaka, Soliton in two-band superconductor, Phys. Rev. Lett. {\bf 88}, 017002 (2001).
\bibitem{Vagov2016} A. Vagov, A. A. Shanenko, M. V. Milo\v{s}evi\'{c}, V. M. Axt, V. M. Vinokur, J. Albino Aguiar, and F. M. Peeters, Superconductivity between standard types: multiband versus single-band materials, Phys. Rev. B {\bf 93}, 174503 (2016).
\bibitem{Salasnich2019} L. Salasnich, A. A. Shanenko, A. Vagov, J. Albino Aguiar, and A. Perali, Screening of pair fluctuations in superconductors with coupled shallow and deep bands: a route to higher-temperature superconductivity, Phys. Rev. B {\bf 100}, 064510 (2019).
\bibitem{Saraiva2020} T. T. Saraiva, P. J. F. Cavalcanti, A. Vagov, A. S. Vasenko, A. Perali, L. Dell’Anna, and A. A. Shanenko, Multiband material with a quasi-1D band as a robust high-temperature superconductor, Phys. Rev. Lett. {\bf 125}, 217003 (2020).
\bibitem{Saraiva2021} T. T. Saraiva, L. I. Baturina, and A. A. Shanenko, Robust superconductivity in quasi-one-dimensional multiband materials, J. Phys. Chem. Lett. {\bf 12}, 11604 (2021).
\bibitem{Shanenko2022} A. A. Shanenko, T. T. Saraiva, A. Vagov, A. S. Vasenko, and A. Perali, Suppression of fluctuations in a two-band superconductor with a quasi-one-dimensional band, Phys. Rev. B {\bf 105}, 214527 (2022).
\bibitem{Moshchalkov2009} V. Moshchalkov, M. Menghini, T. Nishio, Q. H. Chen, A. V. Silhanek, V. H. Dao, L. F. Chibotaru, N. D. Zhigadlo, and J. Karpinski, Type-1.5 superconductivity, Phys. Rev. Lett. {\bf 102}, 117001 (2009).
\bibitem{Kogan2011} V. G. Kogan and J. Schmalian, Ginzburg-Landau theory of two-band superconductors: absence of type-1.5 superconductivity, Phys. Rev. B {\bf 83}, 054515 (2011).
\bibitem{Babaev2012} E. Babaev and M. Silaev, Comment on “Ginzburg-Landau theory of two-band superconductors: absence of type-1.5 superconductivity”, Phys. Rev. B {\bf 86}, 016501 (2012).
\bibitem{Kogan2012} V. G. Kogan and J. Schmalian, Reply to “Comment on ‘Ginzburg-Landau theory of two-band superconductors: absence of type-1.5 superconductivity'", Phys. Rev. B {\bf 86}, 016502 (2012).
\bibitem{Callaghan2005} F.~D. Callaghan, M. Laulajainen, C.~V. Kaiser, and J.~E. Sonier, Field dependence of the vortex core size in a multiband superconductor, Phys. Rev. Lett. \textbf{95}, 197001 (2005).
\bibitem{Fente2016} A. Fente, E. Herrera, I. Guillam\'on, H. Suderow, S. Ma\~nas Valero, M. Galbiati, E. Coronado, and V.~G. Kogan, Field dependence of the vortex core size probed by scanning tunneling microscopy, Phys. Rev. B \textbf{94}, 014517 (2016).
\bibitem{Fente2018} A. Fente, W.~R. Meier, T. Kong, V.~G. Kogan, S.~L. Bud’ko, P.~C. Canfield, I. Guillam\'{o}n, and H. Suderow, Influence of multiband sign-changing superconductivity on vortex cores and vortex pinning in stoichiometric high-Tc~CaKFe$_4$As$_4$, Phys. Rev. B \textbf{97}, 134501 (2018).
\bibitem{Boaknin2003} E. Boaknin, M.~A. Tanatar, J. Paglione, D. Hawthorn, F. Ronning, R.~W. Hill, M. Sutherland, L. Taillefer, J. Sonier, S.~M. Hayden, and J.~W. Brill, Heat conduction in the vortex state of NbSe$_2$: Evidence for multiband superconductivity, Phys. Rev. Lett. \textbf{90}, 117003 (2003).
\bibitem{Gennes1966} P.~G. de~Gennes, \emph {Superconductivity of Metals and Alloys} (Benjamin, New York, 1966).
\bibitem{Gennes1964} P. G. de Gennes, Behavior of dirty superconductors in high magnetic fields, Phys. Kondens. Mater. {\bf 3}, 79 (1964).
\bibitem{Nakai2002} N. Nakai, M. Ichioka, and K. Machida, Field dependence of electronic specific heat in two-band superconductor, J. Phys. Soc. Jpn. {\bf 71}, 23 (2002).
\bibitem{Koshelev2003} A. E. Koshelev and A. A. Golubov, Mixed state of a dirty two-band superconductor: application to MgB$_2$, Phys. Rev. B {\bf 90}, 177002 (2003).
\bibitem{Vargunin2019} A. Vargunin and M. Silaev, Field dependence of the vortex-core size in dirty two-band superconductors, Phys. Rev. B {\bf 100}, 014516 (2019).
\bibitem{Caroli1964} C. Caroli, P.~D. Gennes, and J. Matricon, Bound fermion states on a vortex line in a type-II superconductor, Physics Letters \textbf{9}, 307 (1964).
\bibitem{Gygi1991} F. Gygi and M. Schl\"uter, Self-consistent electronic structure of a vortex line in a type-II superconductor, Phys. Rev. B \textbf{43}, 7609 (1991).
\bibitem{Ichioka2017} M. Ichioka, V.~G. Kogan, and J. Schmalian, Locking of length scales in two-band superconductors, Phys. Rev. B \textbf{95}, 064512 (2017).
\bibitem{Saraiva2017} T.~T. Saraiva, C.~C. de~Souza~Silva, J.~A. Aguiar, and A.~A. Shanenko, Multiband superconductors: disparity between band length scales, Phys. Rev. B \textbf{96}, 134521 (2017).
\bibitem{Chen2020} Y. Chen, H. Zhu, and A.~A. Shanenko, Interplay of Fermi velocities and healing lengths in two-band superconductors, Phys. Rev. B \textbf{101}, 214510 (2020).
\bibitem{Tissen2013} V.~G. Tissen, M.~R. Osorio, J.~P. Brison, N.~M. Nemes, M. Garc{\'{i}}a-Hern\'{a}ndez, L. Cario, P. Rodi\`{e}re, S. Vieira, and H. Suderow, Pressure dependence of superconducting critical temperature and upper critical field of 2H-NbS$_2$, Phys. Rev. B \textbf{87}, 134502 (2013).
\bibitem{Suhl1959} H. Suhl, B. T. Matthias, L. R. Walker, Bardeen-Cooper-Schrieffer theory of superconductivity in the case of overlapping bands, Phys. Rev. Lett. {\bf 3}, 552 (1959).
\bibitem{Moskalenko1959} V. A. Moskalenko, Superconductivity of metals, taking into account the overlapping of energy bands, Phys. Met. Metallogr. {\bf 25}, 8 1959.
\bibitem{Shanenko2011} A.~A. Shanenko, M.~V. Milo\v{s}evi\'{c}, F.~M. Peeters, and A.~V. Vagov, Extended Ginzburg-Landau formalism for two-band superconductors, Phys. Rev. Lett. \textbf{106}, 047005 (2011).
\bibitem{Bogoliubov1970} N. N. Bogoliubov, {\it Lectures on Quantum Statistics, Vol. 2: Quasi-Averages} (Gordon and Breach, New York, 1970). 
\bibitem{Cherny1999} A. Yu. Cherny and A. A. Shanenko, Bound pair states beyond the condensate for Fermi systems below $T_c$ : the pseudogap as a necessary condition, Phys. Rev. B {\bf 60}, 1276 (1999).
\bibitem{Komendova2012} L. Komendov\'{a}, Y. Chen, A.~A. Shanenko, M.~V. Milo\v{s}evi\'{c}, and F.~M. Peeters, Two-band superconductors: hidden criticality deep in the superconducting state, Phys. Rev. Lett. \textbf{108}, 207002 (2012).
\bibitem{Paglione2010} J. Paglione and R. L. Greene, High-temperature superconductivity in iron-based materials.
Nat. Phys. {\bf 6}, 645 (2010). 
\bibitem{Shanenko2015} A. A. Shanenko, J. Albino Aguiar, A. Vagov, M. D. Croitoru, and M. V. Milo\v{s}evi\'{c}, Atomically flat superconducting nanofilms: multiband properties and mean-field theory, Supercond. Sci. Technol. {\bf 28}, 054001 (2015).
\bibitem{Vargas2020} A. A. Vargas-Paredes, A. A. Shanenko, A. Vagov, M. V. Milo\v{s}evi\'{c}, and A. Perali, Crossband versus intraband pairing in superconductors: signatures and consequences of the interplay, Phys. Rev. B {\bf 101}, 094516 (2020).
\bibitem{Hayashi1998} N. Hayashi, T. Isoshima, M. Ichioka, and K. Machida, Low-lying quasiparticle excitations around a vortex core in quantum limit, Phys. Rev. Lett. \textbf{80}, 2921 (1998).
\bibitem{Fetter} A. L. Fetter and J. D. Walecka, {\it Quantum Theory of Many-Particle Systems} (Dover, Mineola, NY, 2003).
\bibitem{Golubov2002} A. A. Golubov, J. Kortus, O. V. Dolgov, O. Jepsen, Y. Kong, O.
K. Andersen, B. J. Gibson, K. Ahn, and R. K. Kremer, Specific heat of MgB$_2$ in a one- and a two-band model from first-principles calculations, J. Phys.: Condens. Matter {\bf 14}, 1353 (2002).
\bibitem{Singh2010} Y. Singh, C. Martin, S. L. Bud’ko, A. Ellern, R. Prozorov, and D. C. Johnston, Multigap superconductivity and Shubnikov–de Haas oscillations in single crystals of the layered boride OsB$_2$ Phys. Rev. B {\bf 82}, 144532 (2010).
\bibitem{Khasanov2010} R. Khasanov, M. Bendele, A. Amato, K. Conder, H. Keller,
H.-H. Klauss, H. Luetkens, and E. Pomjakushina, Evolution of two-gap behavior of the superconductor FeSe$_{1-x}$, Phys. Rev. Lett. {\bf 104}, 087004 (2010).
\bibitem{Kim2011} H. Kim, M. A. Tanatar, Y. J. Song, Y. S. Kwon, and R. Prozorov, Nodeless two-gap superconducting state in single crystals of the stoichiometric iron pnictide LiFeAs, Phys. Rev. B {\bf 83}, 100502(R) (2011).
\bibitem{Bardeen1969} J. Bardeen, R. K\"ummel, A.~E. Jacobs, and L. Tewordt, Structure of vortex lines in pure superconductors, Phys. Rev. \textbf{187}, 556 (1969).
\bibitem{Schmid1966} A. Schmid, A time dependent Ginzburg-Landau equations and its application to the problem of resistivity in the mixed state, Phys. Kond. Materie \textbf{5}, 302 (1966).
\bibitem{Clem1975} J.~R. Clem, Simple model for the vortex core in a type-II superconductor, J. Low Temp. Phys. \textbf{18}, 427 (1975).
\bibitem{Sonier2004} J.~E. Sonier, Investigations of the core structure of magnetic vortices in type-II superconductors using muon spin rotation, Journal of Physics: Condensed Matter \textbf{16}, S4499 (2004).
\bibitem{Tinkham1996} M. Tinkham, \emph { Introduction to Superconductivity} (McGraw-Hill Inc., 1996).
\bibitem{Suderow2005} H. Suderow, V. G. Tissen, J. P. Brison, J. L. Martínez, and S. Vieira, Pressure induced effects on the Fermi surface of superconducting 2H−NbSe$_2$, Phys. Rev. Lett. {\bf 95}, 117006 (2005).
\bibitem{Araujo2009} M. A. N. Ara\'{u}jo, M. Cardoso, and P. D. Sacramento, Single vortex structure in two models of iron pnictide $s^{\pm}$ superconductivity, New J. Phys. {\bf 11}, 113008 (2009).
\end{thebibliography}
\end{document}